\documentclass[reprint,english,aps,pre,floatfix,showpacs]{revtex4-1}

\usepackage{amsmath}
\usepackage{graphicx}
\usepackage{amssymb}
\usepackage{babel}
\usepackage{color}
\usepackage{bm}
\renewcommand{\vec}[1]{\boldsymbol{#1} }

\begin{document}
\title{Superadiabatic forces in Brownian many-body dynamics}

\author{Andrea Fortini$^{\rm 1}$}
\author{Daniel de las Heras$^{\rm 1}$}

\author{Joseph M. Brader$^{\rm 2}$}
\author{Matthias Schmidt$^{\rm 1}$}
\affiliation{$^{\rm 1}$Theoretische Physik II, Physikalisches Institut, Universit\"at Bayreuth, Universit\"atsstra{\ss}e 30, D-95447 Bayreuth, Germany}
\affiliation{$^{\rm 2}$Soft Matter Theory, University of Fribourg, CH-1700 Fribourg, Switzerland}

 


\begin{abstract}

Theoretical approaches to nonequilibrium many-body dynamics generally rest upon an adiabatic assumption, 
whereby the true dynamics is represented as a sequence of equilibrium states. 
Going beyond this simple approximation is a notoriously difficult problem.   
For the case of classical Brownian many-body dynamics we present a simulation method that allows to isolate and 
precisely evaluate superadiabatic correlations and the resulting forces. 
Application of the method to a system of one-dimensional hard particles reveals the importance for the dynamics, as 
well as the complexity, of these nontrivial out-of-equilibrium contributions. 
Our findings help clarify the status of dynamical density functional theory and provide a 
rational basis for the development of improved theories.
\end{abstract}

\maketitle

Adiabatic changes to a dynamical system proceed infinitely slowly and regularly, allowing the system to continuously adapt 
its configuration and remain close to equilibrium. 
When changes occur at finite rates, as is always the case in practice, the dynamics can often be reasonably approximated 
by an adiabatic process. 
This approach has proven very fruitful in treating quantum mechanical problems, from the early work of Ehrenfest~\cite{ehr}, 
Dirac~\cite{dirac}, Born and Fock~\cite{born}, to Berry's discovery of the geometric phase~\cite{berry}. 
In this context, a process may be treated adiabatically when the timescale of the imposed change is 
much larger than the intrinsic timescale of the system.

Out-of-equilibrium systems in classical statistical mechanics can also be treated using adiabatic approximations. 
In contrast to the quantum case, where one typically deals with the full probability distribution, 
adiabatic approximations in classical many-body systems are applied on the coarse-grained level of the correlation functions. 
A well known theory of this type is the dynamical density functional 
theory (DDFT) \cite{evans1979,Tarazona1,Tarazona2}, which predicts the time evolution of the one-body density of Brownian 
particles. 
Within this approach one makes the assumption that the nonequilibrium pair correlations can at any time be 
approximated by those of a fictitious equilibrium system, whose density is given by the instantaneous 
density of the nonequilibrium system \cite{reinhardt}. 
This is equivalent to assuming that the timescale on which the density changes is 
slow compared to the relaxation time of the pair correlations.

 DDFT provides a simple and implementable tool for
 investigating qualitative features of the density evolution.
 Recent applications include the study of active colloidal
 suspensions~\cite{new1}, the modeling of multiple time scales
 during glass formation~\cite{new2}, quasicrystalline order and
 a crystal-liquid state in a soft-core fluid~\cite{new3}, traveling
 crystals in active systems~\cite{new4}, and shock waves in
 capillary collapse of colloids~\cite{new5}. The theory has also
 been recently been generalized to address hydrodynamics
~\cite{new6}. Despite these successes, in many situations the DDFT
 approach either becomes unreliable or breaks down completely. In addition to the well known overestimation
 of relaxation rates~\cite{Tarazona1,Tarazona2,penna2006}, the theory is qualitatively wrong
 for either strongly confined systems or high density states
 around the glass transition.

In this article we address the fundamental limitations of the adiabatic approximation for describing the 
nonequilibrium dynamics of Brownian many-body systems. 
We present a general computer simulation method which enables the superadiabatic contribution to the particle 
motion to be isolated and analyzed in detail. 
As an application, the method is used to study a simple system of confined one-dimensional hard particles.

We consider a system of $N$ interacting Brownian particles. 
The microscopic motion of particle $i$ with position $\vec r_i(t)$  is described by a stochastic differential (Langevin) equation
\begin{equation}
\xi \frac{d {\vec r}_i(t)}{d t} = - \nabla_i U(\vec r^{N}\!\!,t)+\vec  X_{i}(t) \ ,
\label{BD}
\end{equation}
where $\xi$ is the friction coefficient, $U(\vec r^{N}\!\!,t)$ is the potential energy of configuration 
$\vec r^{N}\!$,  $\vec  X_{i}(t)$ is a Gaussian random force,  and  $\nabla_i$ indicates
  the partial derivative with respect to the position of particle $i$.

The evolution of the probability distribution, $P(\vec r^{N}\!\!,t)$, is given by the Smoluchowski 
equation \cite{dhont}
\begin{equation}
\xi \frac{\partial P(\vec r^{N}\!\!,t)}{\partial t} = \sum_{i=1}^{N} \nabla_{i} 
\cdot \left[ k_{B} T \nabla_{i} +\nabla_{i} U(\vec r^{N}\!\!,t)\right]P(\vec r^{N}\!\!,t) \ ,
\label{SM}
\end{equation}
where $k_{B}$ is the Boltzmann constant and $T$ is the temperature.
We restrict our discussion to potential energies of the form
$U(\vec r^{N}\!\!,t)= \sum_i V_{\rm ext}(\vec r_i) +\sum_{i<j} \phi(|\vec r_{i}-\vec r_{j}|)$,
where $V_{\rm ext}$ and $\phi$ are the external and pair potentials, respectively.  
Integrating the probability distribution yields the one-body density
\begin{equation}
\rho^{(1)}(\vec r_1,t)=N\!\int \vec d r_{2}  \dots \int d \vec r_{N} P(\vec r^{N}\!\!,t),
\label{rho1}
\end{equation}
and two-body density 
\begin{equation}
\rho^{(2)}(\vec r_{1},\vec r_{2},t)=N(N-1)\! \int d \vec r_{3}  \dots \int d \vec r_{N} P(\vec r^{N}\!\!,t),
\label{rho2}
\end{equation}
which provide a coarse-grained description of the instantaneous microstructure~\cite{Hansen:2013uv}. 

Integration of Eq.~(\ref{SM}) over all but one of the coordinates yields a continuity 
equation for the density~\cite{Archer:2004eq} 
\begin{equation}
\frac{\partial \rho^{(1)}(\vec r,t)}{\partial  t} = -\nabla\cdot {\bf J}(\vec r, t).
\label{SM1}
\end{equation}
The one-body current is related to the total force according to 
${\bf J}(\vec r,t)=\xi^{-1}\rho^{(1)}(\vec r,t)\,{\bf F}(\vec r,t)$, where 
\begin{equation}
{\bf F}(\vec r,t)\!=\! -k_BT \nabla \ln\left( \rho^{(1)}(\vec r,t)\right) - \nabla V_{\rm ext}(\vec r, t)  
+ {\bf F}_{\rm int}(\vec r, t).
\label{SM1}
\end{equation}
The internal force acting on the density field at point $\vec r$ arises 
from pair interactions and is defined by ${\bf F}_{\rm int}(\vec r, t)={\bf I}(\vec r, t)/\rho^{(1)}(\vec r,t)$, 
with the exact force integral given by
\begin{equation}
{\bf I}(\vec r, t)=-\int d \vec r'\rho^{(2)}(\vec r , \vec r', t)  \nabla' \phi(|\vec r-\vec r'|). 
\label{int1}
\end{equation}
Equations \eqref{SM1}-\eqref{int1} represent the first in a nonequilibrium hierarchy of equations  for the 
$n$-point density functions \cite{Hansen:2013uv}. Evaluation of the nonequilibrium pair density and the 
force integral \eqref{int1} constitutes the primary aim of this article. 

In order to isolate the physical processes of interest we split the force integral into 
adiabatic and superadiabatic contributions, 
$\vec I(\vec r,t)=\vec I_{\rm ad}(\vec r,t)+\vec I_{\rm sad}(\vec r,t)$,
where
\begin{align}
{\bf I}_{\rm ad}(\vec r,t)&= -\int  d \vec r' \rho^{(2)}_{\rm ad}(\vec r , \vec r',t)  \nabla' \phi(|\vec r - \vec r'|), 
\label{iad}
\\
{\bf I}_{\rm sad}(\vec r,t)&=-\int d \vec r' \rho^{(2)}_{\rm sad}(\vec r , \vec r',t)  \nabla' \phi(|\vec r - \vec r'|).
\label{isad}
\end{align}
The adiabatic two-body density, $\rho^{(2)}_{\rm ad}(\vec r , \vec r',t)$, used to evaluate \eqref{iad} is that of an 
equilibrium system with one-body density $\rho^{(1)}(\vec r,t)$. 
The DDFT employs equilibrium methods to obtain an approximation to \eqref{iad} and thus 
implicitly uses the approximation ${\bf I}(\vec r,t) = {\bf I}_{\rm ad}(\vec r,t)$.

Here we propose a simple and general computational scheme to 
analyze the superadiabatic contribution to the dynamics. 
The method is implemented as follows:  

\begin{itemize}
\item Sample the two-body density $\rho^{(2)}(\vec r,\vec r',t_{s}) $ and one body density $\rho^{(1)}(\vec r,t_{s})$ with 
nonequilibrium computer simulations at a time $t=t_{s}$.

\item Calculate the force integral ${\bf I}(\vec r,t_s)$ via Eq.(\ref{int1}).

\item Find a fictitious external potential $V_{\rm ad}(\vec r,t_s)$ (henceforth referred to as the adiabatic potential) 
that generates in an equilibrium simulation the instantaneous nonequilibrium density 
$\rho^{(1)}(\vec r,t_{s})\equiv\rho_{\rm ad}^{(1)}(\vec r,t_s)$.

\item Perform an equilibrium simulation with the newly found adiabatic potential and sample 
$\rho^{(2)}_{\rm ad}(\vec r , \vec r',t_s)$ 
  via Eq.  \eqref{rho2} with the adiabatic probability distribution $P_{\rm ad}(\vec r^N)$ that possesses
    the equilibrium form, and is hence defined as the (normalized)
    Boltzmann factor of the internal interactions
    and the external potential energy $V_{\rm ad}(\vec r,t_s)$.    

\item  Calculate the force integral ${\bf I}_{\rm ad}(\vec r,t_s)$ using (\ref{iad}).

\item Identify the superadiabatic force integral by computing the difference 
${\bf I}_{\rm sad}(\vec r,t_s)={\bf I}(\vec r,t_s)-{\bf I}_{\rm ad}(\vec r,t_s)$.

\end{itemize}

\begin{figure}
\includegraphics[width=7.7cm]{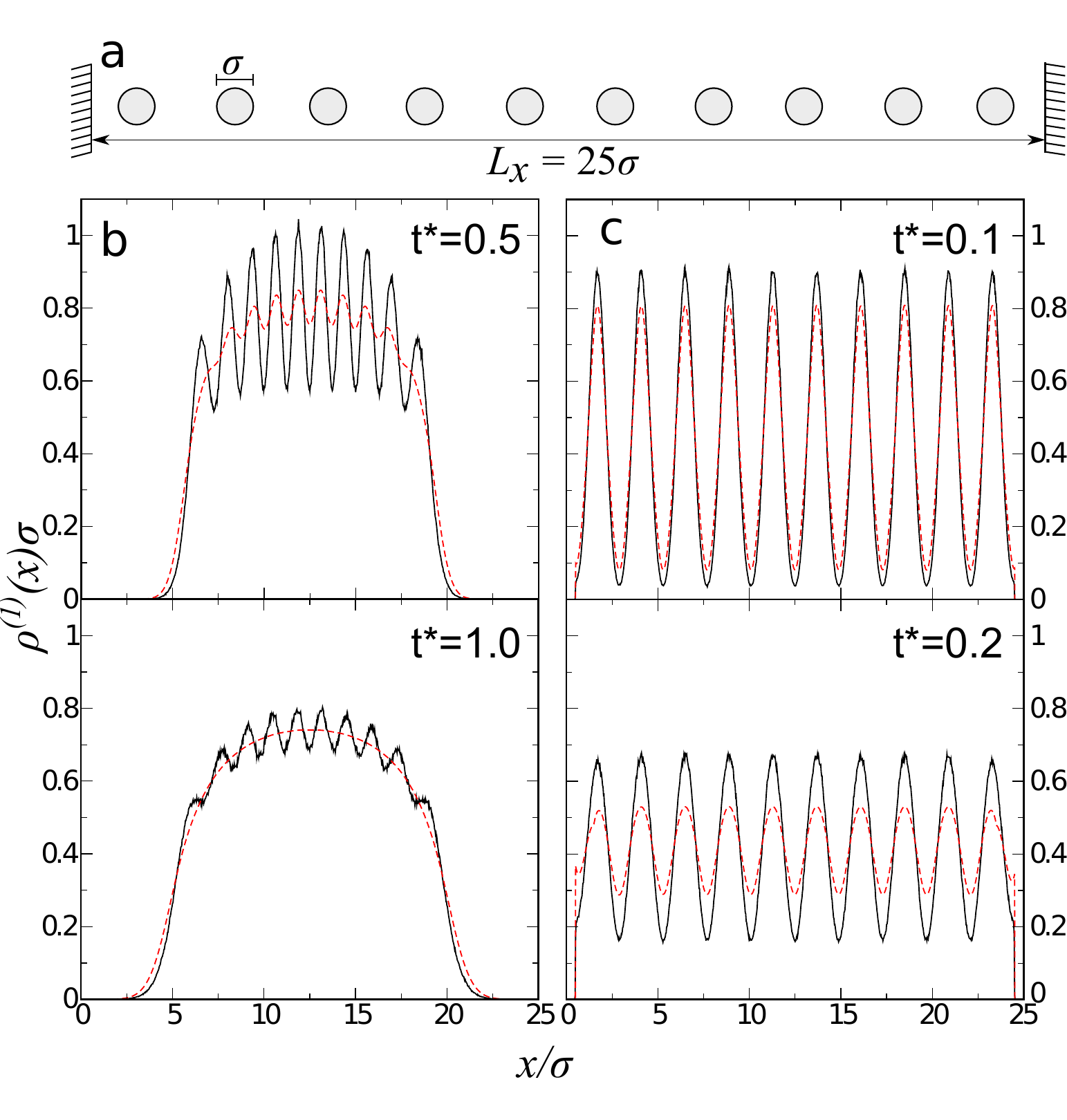}
\caption{(Color online) a) Schematic representation of a system of $10$ hard particles confined between hard walls. b) Density profiles calculated from BD simulations (black-solid line) and DDFT (red-dashed line) at reduced time $t^*=t_s/\tau_B$=0.5 (top), and 1.0 (bottom) for a system initialized in a parabolic trap. c) Density profiles calculated from BD simulations (black-solid line) and DDFT (red-dashed line) at reduced time $t^*$=0.1 (top), and 0.2 (bottom)  for a system initialized in a crystal structure  (for a comparison of the adiabatic forces see  appendix B).
 }
\label{fig1}
\end{figure}

To provide a concrete implementation of this scheme we consider a one-dimensional system (schematically  shown in Fig.~\ref{fig1}a) of $N=10$ quasi-hard particles of  length $\sigma$ confined between quasi-hard walls with separation distance $L_{x}$. 
Our choice to investigate hard particles is motivated by the fact that this system, although simple, is sufficient to demonstrate 
both the utility of our computational approach and the limitations of the adiabatic approximation. 
Higher dimensional models can also be addressed with our method.

The 
pair interaction between particles at distance $r$  is  $\phi(r)/k_B T = (\sigma/r)^{42}$  for $r<  \sigma $ and vanishes otherwise.
The particle-wall interaction potential is $V_{\rm ext}(x)/k_B T =(\sigma/x)^{42}$ if the wall-particle distance $x<\sigma/2$ and vanishes otherwise.
The exponent 42 has been chosen because it provides a good balance between the need of  a steep repulsive potential, which represents the hard-core of the particles, and computation efficiency.

 \begin{figure}
\vspace*{-0.6cm}
\includegraphics[width=8.cm]{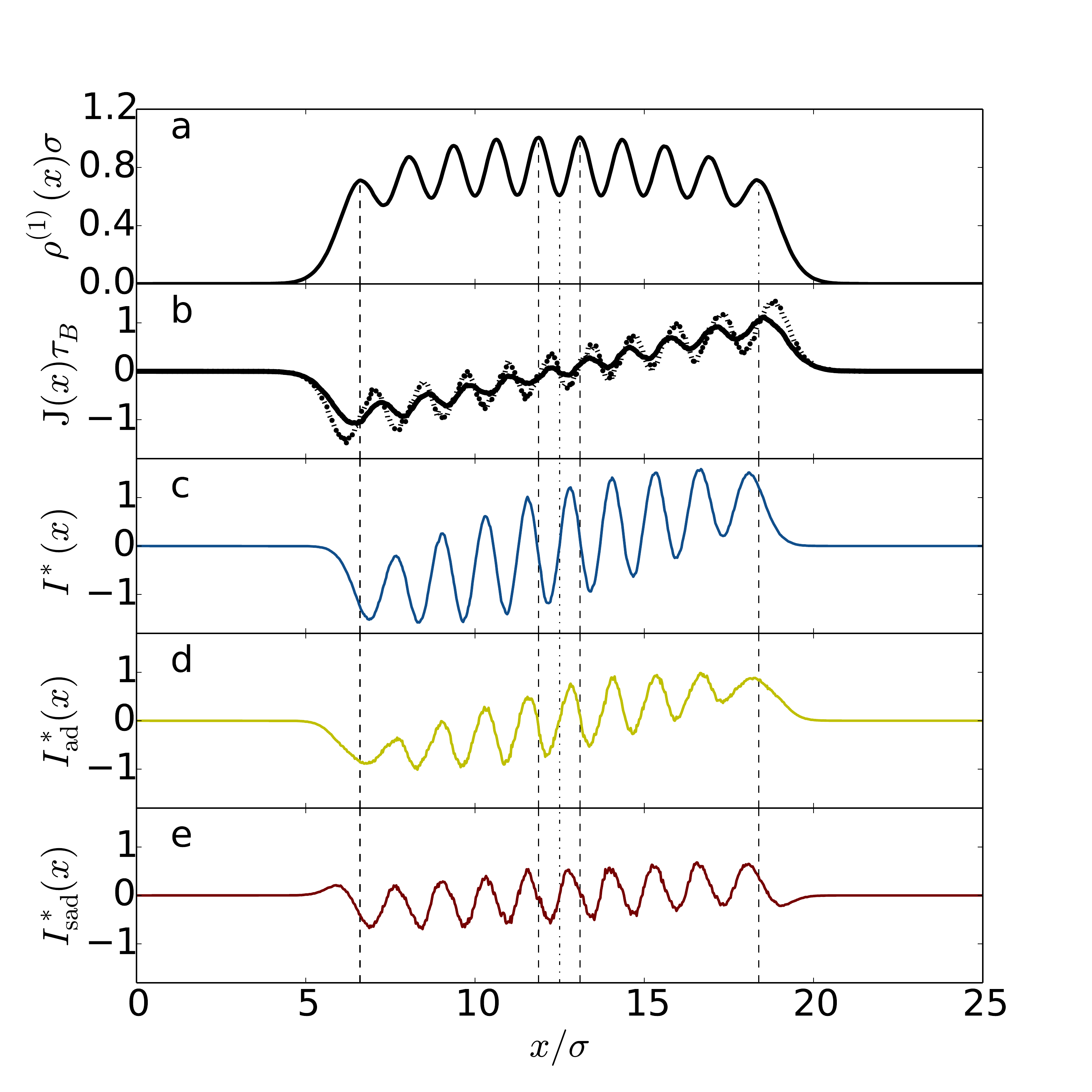}
\vspace*{-0.2cm}
\caption{(Color online) System relaxing following release from a parabolic trap ($\!t_{s}\!=\!0.5 \tau_{B}$). 
a) One-body density $\rho^{(1)}(x,t_{s})$. 
b) Total current ${\rm J}(x)$ from BD  (continuos line) and DDFT (dotted line).
c) Total force integral $I^*(x)=I(x) \sigma^{2}/k_{B} T$.  
d) Adiabatic force $I^*_{\rm ad}(x)=I_{\rm ad}(x) \sigma^{2}/k_{B} T$. 
e) Superadiabatic force $I^*_{\rm sad}(x)=I_{\rm sad}(x) \sigma^{2}/k_{B} T$.
The vertical dashed lines  serve as a guide for the eye. }
\label{parab}
\end{figure}

In order to explore some typical nonequilibrium situations, we initialize the system in two distinct states. 
Firstly,  the system is equilibrated in a parabolic trap, $V=\alpha x^{2}$, with $\alpha=10 k_B T/\sigma^2$, which is then   suddenly removed at time $t=t_{\rm eq}$. 
Secondly, we initialize the particles in a `crystal' structure, i.e. the particles are placed on an ordered lattice between the two walls. In both cases we follow the free relaxation of the system for $t>t_{\rm eq}$.

We first compare the time evolution of the density obtained from 
Brownian dynamics (BD) simulation~\cite{Ermak:1975un,Allen1987} with that predicted by DDFT (see the appendix A and B for 
details of the DDFT implementation), for a system with a reduced average density $\rho\sigma=0.4$. 
 Figures \ref{fig1}b and \ref{fig1}c show the density profiles obtained at two different times for parabolic trap and crystal initial conditions, respectively.  
Before removal of the trap we 
allow an equilibration time $t_{\text{eq}}= 5 \tau_{B}$, with Brownian 
time $\tau_B=\sigma^2\xi/(k_B T)$, and then follow the evolution of the system until time $t-t_{\text{eq}}=t_{s}$ ($t_{\text{eq}}=0$ 
for crystal initial conditions). 

Although BD and DDFT show a similar trend, it is clear that the damping of the oscillations proceeds more slowly in the simulation. 
This well-known discrepancy \cite{Tarazona1,Tarazona2} is commonly attributed to ensemble differences: 
canonical in BD and grand canonical in DDFT (Note that the development of canonical equilibrium DFT is a question of current research~\cite{can1,can2}, and that a canonical version of DDFT is not available at present). 
However, in the limit $t_s\rightarrow\infty$ both profiles 
become very similar (see appendix) indicating 
that for this number of particles ensemble differences do not account for the discrepancy between BD and DDFT. 
We have performed the same analysis 
up to $N=30$ and found similar discrepancies. 
As we will demonstrate below, the failure of DDFT in adequately describing the dynamics lies in the neglect of 
superadiabatic forces. 

 \begin{figure}
\vspace*{-0.5cm}
\includegraphics[width=8cm]{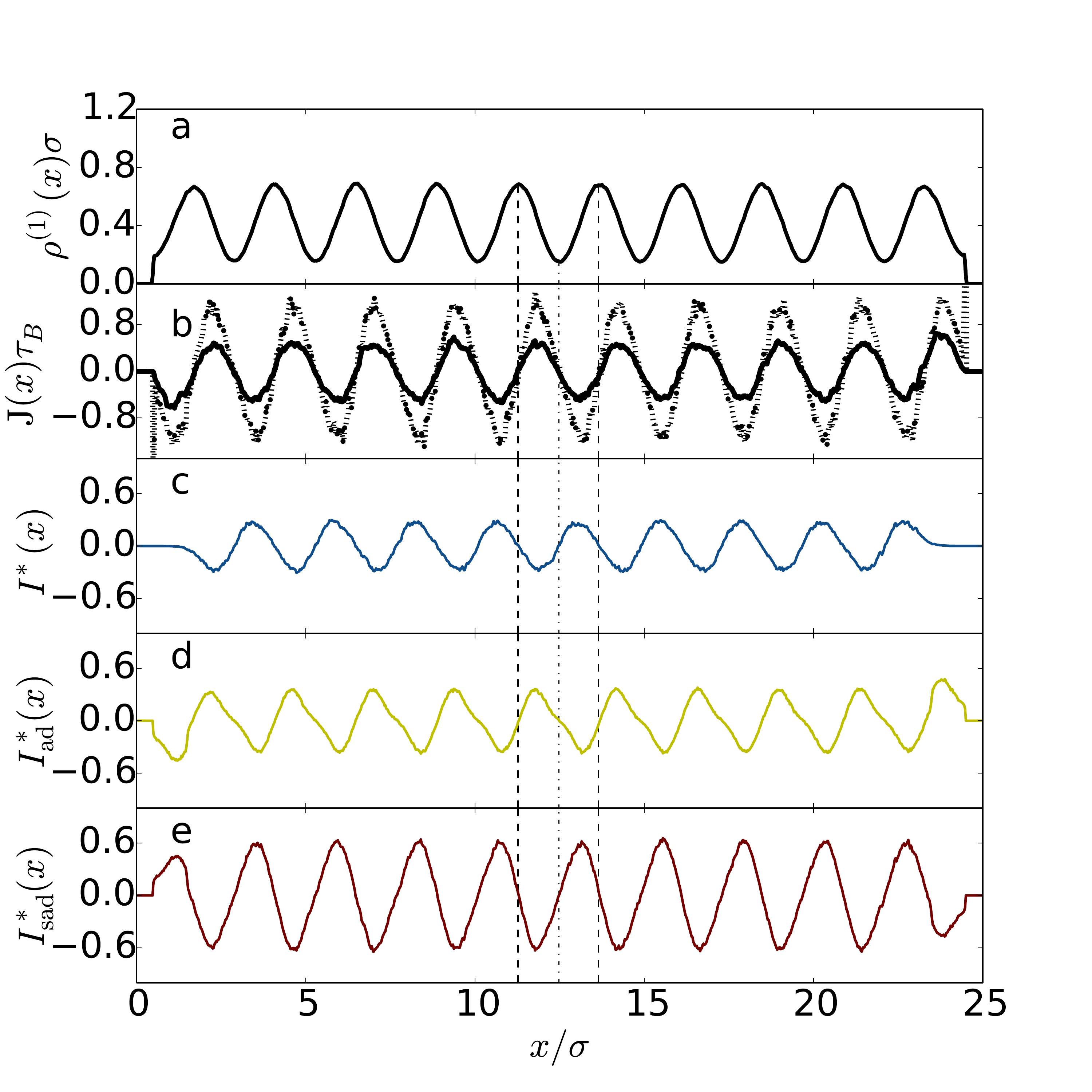}
\caption{(Color online) Same as Fig.~\ref{parab} but for the system initialized in a crystal state and sampled at time $t_{s}=0.2 \tau_{B}$.
}
\label{crys}
\end{figure}

To implement our numerical scheme we discretized the box length $L_{x}$ into bins of width $d_x\!=\!0.0025\sigma$.
After equilibration both $\rho^{(1)}_{BD}(x_l,t_s)$ and $\rho^{(2)}_{BD}(x_l,x_k,t_s)$ 
are sampled at time $t_s$ by averaging over $M=10^6$ independent trajectories, where $l,k$ are indexes that run over all discrete bins. 
The search for the adiabatic potential is then carried out using a series of canonical Monte Carlo (MC) 
simulations~\cite{Frenkel}: we discretize $V_{\rm ad}(x_l)$, initializing it to an arbitrary 
function~\footnote{In order to speed up convergence of the iteration procedure, a physical estimation of the potential can be obtained by 
inversion of the  target density $\rho^{(1)}_{BD}$ with density functional theory.}, and perform a MC simulation 
with this external potential. Following 1000 equilibration steps the one-body density is sampled for 1000 steps.
The computed $\rho^{(1)}(x_l)$  is then compared in each bin with the desired target density $\rho^{(1)}_{BD}(x_l)$.  
If $\rho^{(1)}(x_l)>\rho^{(1)}_{BD}(x_l)$ then the potential in bin $l$ is increased, otherwise it is decreased. 
This process is iterated until $|\rho^{(1)}_{BD}(x_l,t_s)-\rho^{(1)}(x_l)|<0.005 \sigma^{-1}$. 
The adiabatic potential thus obtained is then used in a final, longer MC simulation: 
After $10^5$ equilibration steps the (adiabatic) one- and two-body densities are sampled for $10^5$ 
steps and the superadiabatic force is computed 
using \eqref{isad}.

Figure~\ref{parab}a shows the equilibrium density 
at $t_s=0.5 \tau_b$ for the system initialized  in a parabolic trap  with average density 
$\rho\sigma=0.4$. 
Clearly, the particles have not reached the 
walls located at $x/\sigma\!=\!0$ and $x/\sigma\!=\!25$. 
Figure~\ref{parab}b compares the current from BD (see the appendix C for simulation details) with 
that obtained by using the simulated $\rho^{(1)}(\vec r,t)$ as input to the DDFT. 
The DDFT current is larger in magnitude than the simulated one. 
This is compatible with various observations that DDFT dynamics
are faster than in simulation.

The expanding set of particles is characterized by a pair force $I(x)$ shown in Fig.~\ref{parab}c, which corresponds to the pair force acting at position $x$.
The overall slope of the curve indicates that the system experiences a force pushing the density outwards towards the walls. 
On top of this overall expansion force is superposed a local oscillatory structure indicating 
that each particle (except the first and last) is subject to a confining force arising from the cage of nearest neighbors.

 \begin{figure}
\includegraphics[width=8cm]{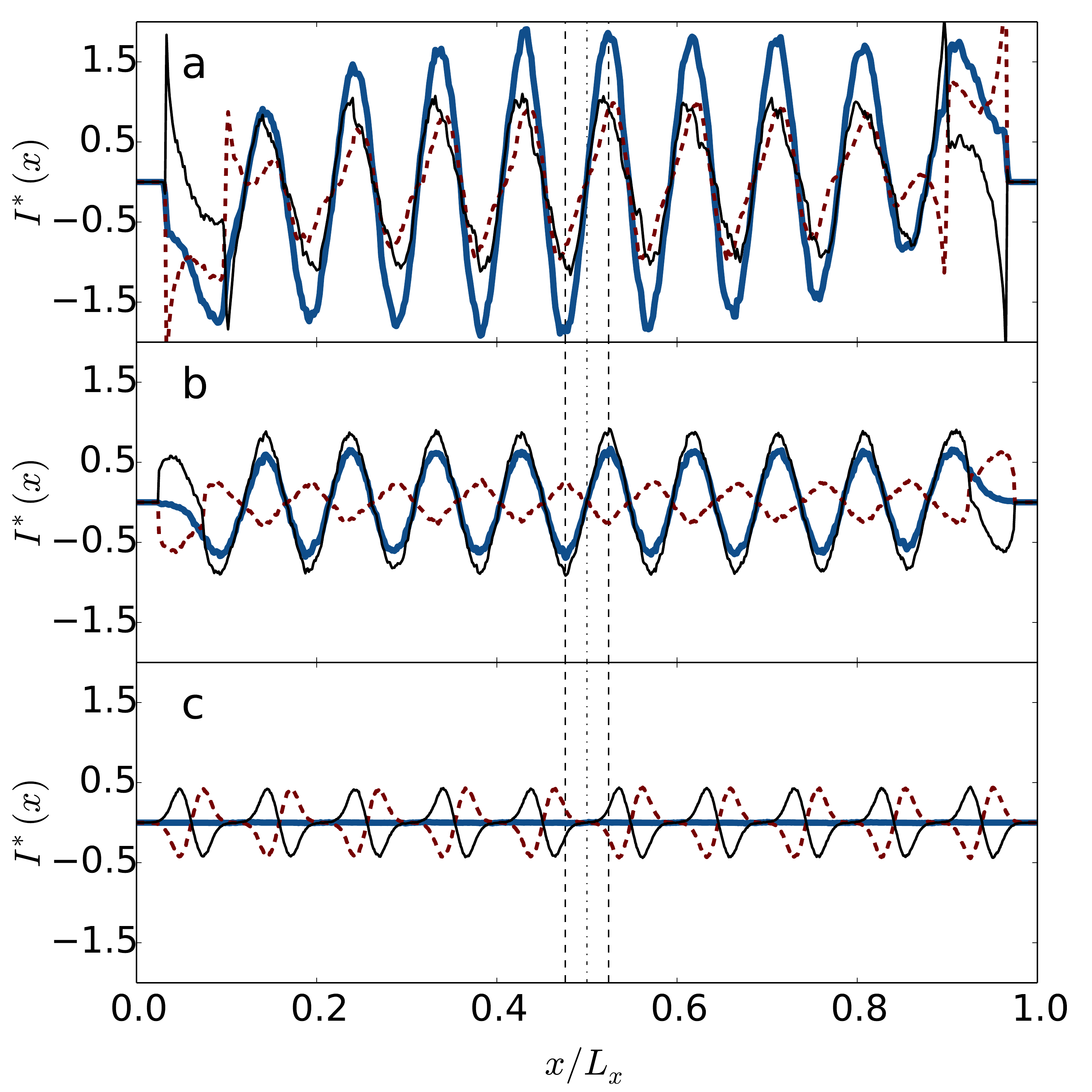}
\caption{(Color online) Total  $I^{*}(x)=I(x) \sigma^{2}/k_{B} T$ (thick line),  adiabatic  $I^{*}_{\rm ad}(x)=I_{\rm ad}(x) \sigma^{2}/k_{B} T$ (dashed line) and superadiabatic  $I^{*}_{\rm sad}(x)=I_{\rm sad}(x) \sigma^{2}/k_{B} T$ (thin line) force integrals at different densities for the system initialized in a crystal state and  sampled at time $t_{s}=0.2 \tau_{B}$. The vertical dashed lines  serve as a guide for the eye. a) $\rho \sigma=0.25$ ($L_x/\sigma=40 $). b) $\rho \sigma=0.5$ ($L_x/\sigma=20$). c) $\rho \sigma=0.67$ ($L_x/\sigma=15$).
}
\label{dens}
\end{figure}

Figure~\ref{parab}d shows the adiabatic contribution $I_{\rm ad}$.  
Not only is the functional form very similar to that of $I$, but the overall slope is captured almost 
completely by $I_{\rm ad}$. 
The superadiabatic contribution, $I_{\rm sad}$, shown in Fig.~\ref{parab}e is of 
roughly the same magnitude as $I_{\rm ad}$ and has a similar global slope.
The total, adiabatic and superadiabatic forces all oscillate in phase, suggesting that the influence of the 
superadiabatic contribution to the dynamics could be approximated by a 
global scaling of $I_{\rm ad}$.

However, the picture becomes more complex when considering a crystal initial state.  
Figure~\ref{crys}a shows the 
density profile at $t_s=0.2 \tau_b$ for the system with average density 
$\rho\sigma=0.4$.  
For this relatively short time the equidistant density peaks are rather symmetric about 
their initial positions. 
As shown in Fig.~\ref{crys}b, the magnitude of the simulated current is also smaller than that from 
DDFT for this initial condition. 

The full force integral $I$, shown in figure~\ref{crys}c, has no global slope
, as there is no tendency for the system to expand.
Moreover, the oscillatory structure of the peaks in $I$ is precisely what one would expect as a result of confinement 
by repulsive neighboring particles. 
Most interestingly, the adiabatic contribution (Fig.~\ref{crys}d) oscillates \emph{out of phase} with respect to $I$; 
the adiabatic approximation is qualitatively wrong 
leading to a large superadiabatic correction (Fig.~\ref{crys}e). 
The adiabatic forces acting on the density field erroneously predict that repulsive 
interactions with neighbors should enhance the relaxation rate. 
 Note that the ideal diffusion 
creates the primary contribution. Any interaction
     contributions generally subtract from this and tend to
     slow down the dynamics. The subtraction that the adiabatic piece
     generates is, however, too small or even has the wrong
     sign. This explains, e.g., that in Fig.~\ref{parab} $I^*$ must be larger in
     magnitude than $I^*_{\rm ad}$.
It is thus clear that superadiabatic forces cannot in general be 
accounted for by rescaling the adiabatic 
forces, but rather represent a distinct additive contribution, which is essential to recover the correct physical 
behavior of the system. 
This finding is 
consistent with the recently developed power functional theory \cite{Schmidt:2013kn}, which 
is an exact generalization of equilibrium density functional theory to nonequilibrium Brownian dynamics. 
Within the power functional approach the superadiabatic forces are generated by an additive 
contribution to the power dissipation functional. 

The 
unexpected behavior of the adiabatic force can be elucidated by a systematic 
investigation of the system with crystal initial conditions at various densities. 
The total, adiabatic and superadiabatic forces are shown in Fig.\ 4 at
different densities. For the highest value considered, $\rho\sigma =
0.67$ (Fig.\ 4a), the adiabatic force integral oscillates in phase
with the total force integral. However, as the density is decreased we
find very different behavior. The functional form of the forces at
density $\rho\sigma = 0.5$ (Fig.\ 4b) is very similar to the one shown
in Fig.\ 3d for $\rho\sigma = 0.4$, i.e.\ the adiabatic force
oscillates out-of-phase with respect to the total pair force. We can
thus conclude that the behavior of the adiabatic force and
consequently the extent and importance of the superadiabatic
contribution depend nontrivially on the average distance between the
particles (the time evolution of the superadiabatic force is analyzed briefly in appendix  D).  
The out-of-phase behavior at low density can be elucidated
by analyzing the forces	at density $\rho\sigma = 0.25$ (Fig.\
4c). Here the total force vanishes, indicating that no pair
interactions have occurred at the sampling time. 
Nevertheless, the adiabatic force is 
non-zero and, instead of a confining force, we find a force that moves particles away from the
peak centers. 
Since the adiabatic force is an equilibrium
contribution, it includes all possible configurations for a given
external potential. Among those are configurations with two particles inside the same potential well, 
which give rise to an erroneous repulsive force.
Since the total force vanishes, the
superadiabatic and adiabatic contributions are exactly opposite. 
At higher densities, the density peaks are narrower,  
configurations with two particles in the same peak are less likely to
occur and the adiabatic force has only confining contributions due to interactions with neighbor peaks.

In conclusion we have developed a general method for 
estimating
superadiabatic forces in a system of interacting 
Brownian particles. 
We have applied the method to confined
hard particles in one dimension and thus revealed two important features 
of the superadiabatic force: (i) It is of the same magnitude as the adiabatic forces and thus 
cannot be regarded as a small correction, 
(ii) The behavior depends in a nontrivial way on the average distance between the particles. 
These findings suggest that the validity of the adiabatic approximation depends sensitively upon the 
particular dynamic path taken by the system as it relaxes through the space of density functions.

Although we have applied our method to BD simulation data, we note that it could,
in principle, also be used to determine the superadiabatic contribution in colloidal experiments. 
An adiabatic external field could be 
obtained for example by application of a light field~\cite{Bechinger:2000jm}.

\appendix
\section{Density Functional Theory}
The exact density functional of a one dimensional system of hard-particles was developed by Percus \cite{Percus1976}. The free energy is
\begin{equation}
F[\rho^{(1)}]=F_{\text{id}}[\rho^{(1)}]+F{_\text{ex}}[\rho^{(1)}],
\end{equation}
where $F_{\text{id}}$ is the ideal gas contribution and the excess part $F{_\text{ex}}$ accounts for the excluded-volume interactions between the particles:
\begin{eqnarray}
&&\beta F_{\text{id}}[\rho^{(1)}] = \int dx\rho^{(1)}(x)\left(\ln(\Lambda\rho^{(1)}(x))-1\right),\\
&&\beta F_{\text{ex}}[\rho^{(1)}] =  \nonumber \\
&&-\frac12\int dx\left(\rho^{(1)}(x-\sigma/2)+\rho^{(1)}(x+\sigma/2)\right)\ln(1-\eta(x)). \nonumber
\end{eqnarray}
In the above expressions $\beta=1/k_{B}T$ with $k_{B}$ the Boltzmann constant and $T$ the temperature. $\Lambda$ is the (irrelevant) thermal wavelength, $x$ is the space coordinate, and $\eta(x)$ is the local packing fraction, defined as
\begin{equation}
\eta(x)=\int_{x-\sigma/2}^{x+\sigma/2}dx'\rho^{(1)}(x'),
\end{equation} 
with $\sigma$ the particle length.

The grand canonical density functional is
\begin{equation}
\beta\Omega[\rho^{(1)}]=F[\rho^{(1)}]+\int dx\rho^{(1)}(x)(V_{\text{ext}}(x)-\mu),
\end{equation}
where $\mu$ is the chemical potential and $V_{\text{ext}}$ is the external potential. 

\begin{figure}[htdp]
\includegraphics[width=7cm]{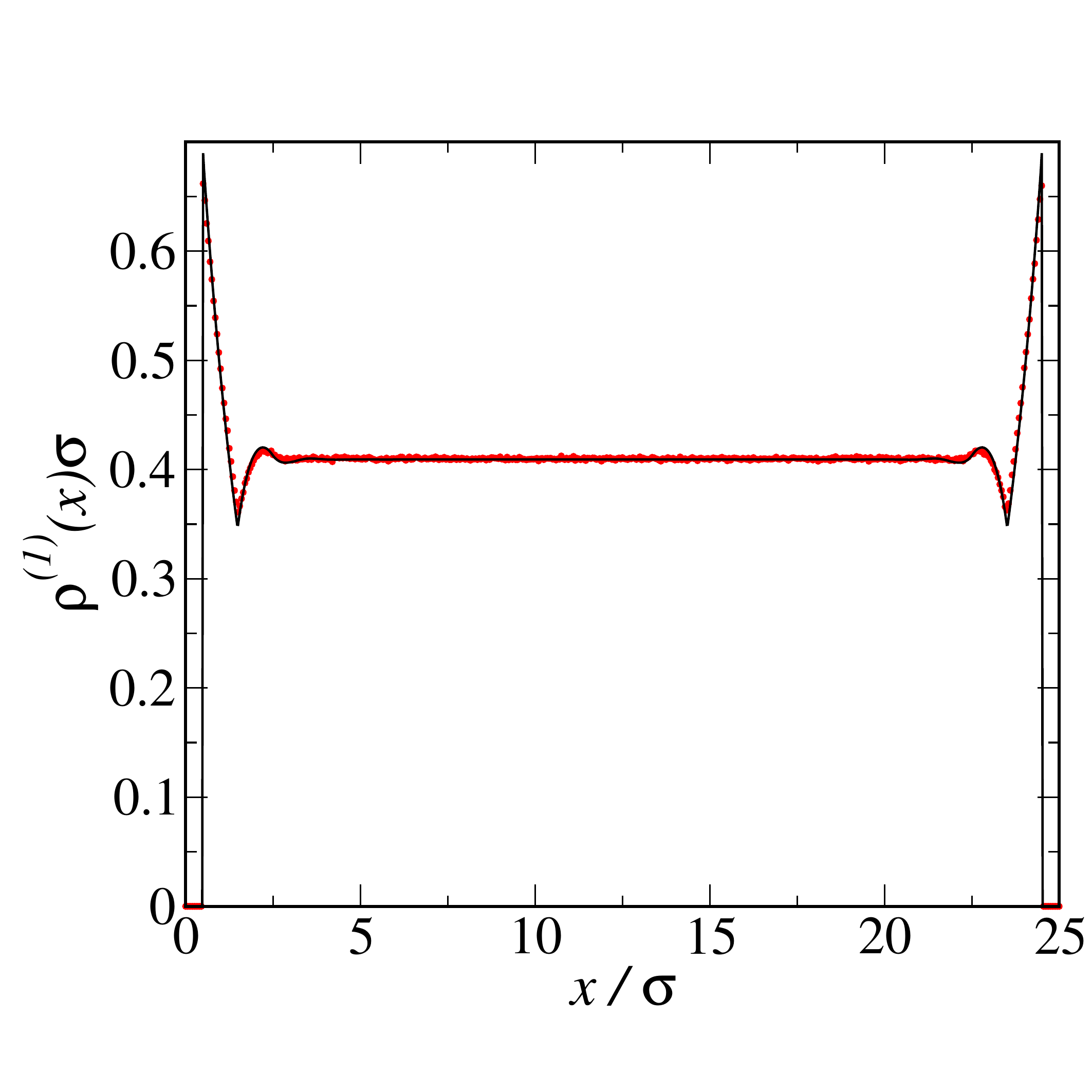}
\caption{Equilibrium density profiles of a system of hard particles confined between hard walls separated by a distance $25\sigma$. Black-solid line: the grand canonical density profile obtained with DFT at a chemical potential $\beta\mu=0.3258$ that corresponds to an average number of particles $\langle N \rangle=10$. Red circles: canonical MC simulation of a system of $N=10$ particles.}
\label{fig_s1}
\end{figure}

The equilibrium density profiles are those that minimize the grand potential density functional at constant $\mu$. We use a standard conjugated gradient method to minimize $\Omega$.  In order to compare the results with the canonical Brownian dynamics (BD) or Monte Carlo (MC) simulation we find the chemical potential for which the average number of particles is equal to the number of particles in the simulation. Given the reduced number of particles the canonical and the grand canonical ensembles are not equivalent. The grand canonical density profiles are combinations of canonical profiles. We show in Fig. \ref{fig_s1} the equilibrium density profiles of a system of $N=10$ particles confined in a pore with $L_x=25\sigma$ in the canonical (MC) and grand  canonical (DFT) ensembles. The differences are small and do not justify the large discrepancy between the predictions of Dynamic Density Functional Theory (DDFT) and BD. 

\section{Dynamic Density Functional Theory}
In DDFT the time evolution of the density profile is governed by the continuity equation \cite{Tarazona1,Tarazona2}
\begin{equation}
\frac{\partial\rho^{(1)}(\vec{r},t)}{\partial t}=-\nabla\cdot {\bf J}_{\rm ad}(\vec{r},t),
\end{equation}
where $\vec r$ is the coordinates vector, $t$ is the time, and $\vec J_{\rm ad}$ is the adiabatic current given by
\begin{equation}
\xi {\bf J}_{\rm ad}(\vec{r},t)=-\rho^{(1)}(\vec r,t)\left(\nabla\frac{\delta F[\rho^{(1)}]}{\delta\rho^{(1)}(\vec r,t)}+\nabla V_{\rm ext}(\vec r,t)\right),
\end{equation}
where $\xi$ is the friction coefficient and $V_{ext}$ is an external potential.

For the one-dimensional system of particles analysed here, the equation for the time evolution of the density according to DDFT reads
\begin{eqnarray}
&&\xi\frac{\partial\rho^{(1)}(x,t)}{\partial t} = \frac{\partial^2\rho^{(1)}(x,t)}{\partial x^2}  +\nonumber \\
&+&\frac{\partial}{\partial x}\left[\rho^{(1)}(x,t)\left(\frac{\rho^{(1)}(x+\sigma,t)}{1-\eta(x+\sigma/2)}\right.\right.
-\left.\left.\frac{\rho^{(1)}(x-\sigma,t)}{1-\eta(x-\sigma/2)}\right)\right] \nonumber\\
&+&\frac{\partial}{\partial x}\left({\rho^{(1)}(x,t)\frac{\partial V_{\rm ext}(x,t)}{\partial x}}\right).
\end{eqnarray}

\begin{figure}
\includegraphics[width=7cm]{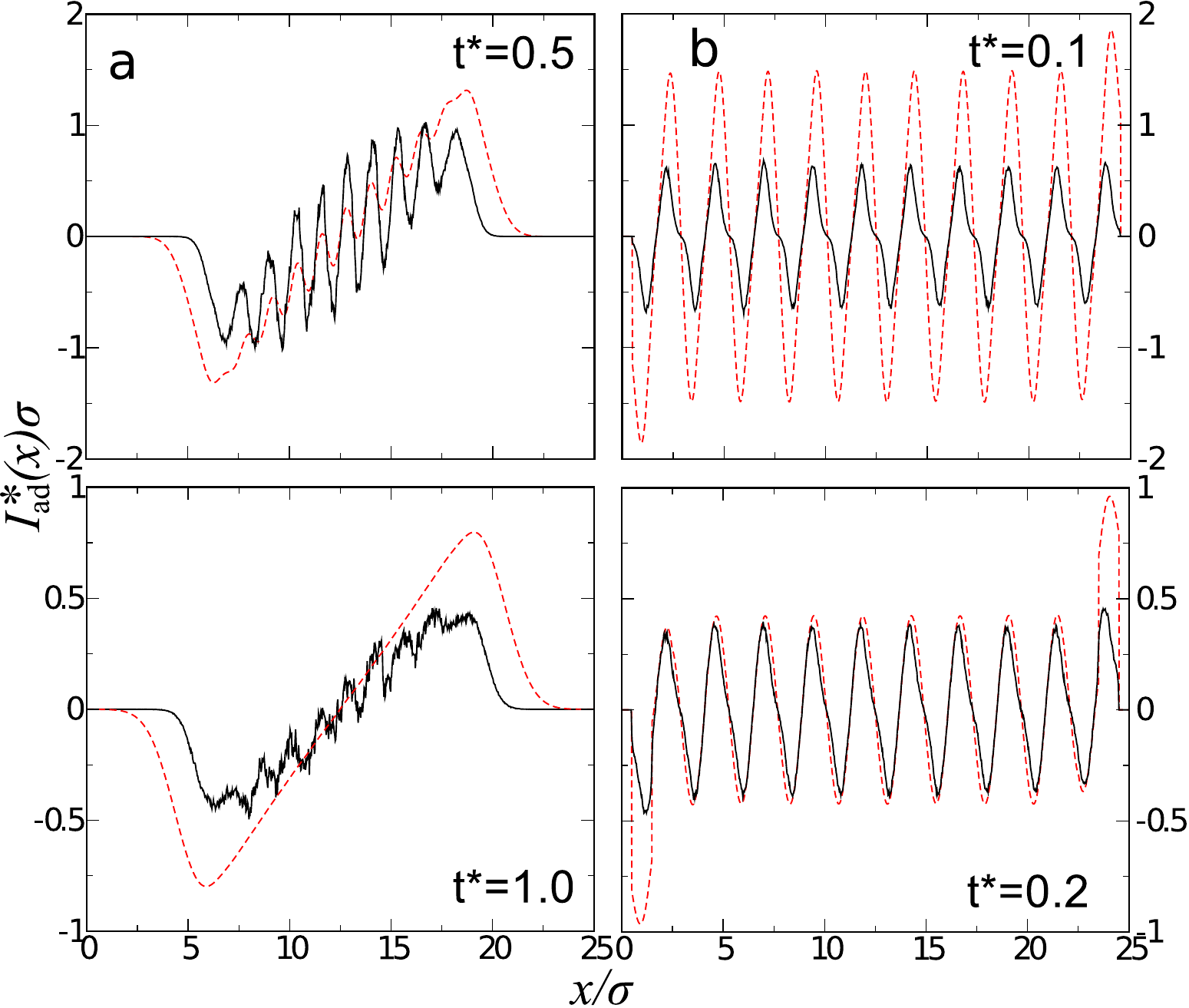}
\caption{a) Adiabatic force calculated from BD simulations (black-solid line) and DDFT (red-dashed line) at reduced time $t^*=t_s/\tau_B$=0.5 (top), and 1.0 (bottom) for a system initialized in a parabolic trap. b) Adiabatic force calculated from BD simulations (black-solid line) and DDFT (red-dashed line) at reduced time $t^*$=0.1 (top), and 0.2 (bottom)  for a system initialized in a crystal structure.}
\label{fig2}
\end{figure}

The comparison between simulation and DDFT results for the density  are shown in Fig. 1 of the the main article. 
Figure~\ref{fig2}a,b show the same comparison comparison for the computed adiabatic contribution. 

\section{Measurements of the Current in Brownian Dynamics Simulations}
In order to measure the current in Brownian dynamics simulations we solve the one dimensional  continuity equation\begin{equation}
\frac{\partial \rho^{(1)}( x,t)}{\partial  t}=-\frac{\partial {\rm J}_{x}(x,t)}{\partial  x} .
\end{equation}

The average 
$$\langle \frac{\partial \rho^{(1)}(x,t_{s})}{\partial  t} \rangle \simeq \langle \frac{\Delta \rho^{(1)}(x,t_{s})}{\Delta  t}\rangle $$ 
is computed over $10^{6}$ independent trajectories at a fixed time $t_{s}$. 

In order to carry out the calculation, we divide the one dimensional simulation box in bins of length $x_{bin}$ and accumulate the histogram of the local density changes  
$$\Delta \rho^{(1)}(x,t_{s})=\frac{n(x,t_{s})-n(x,t_{s}-\Delta t)}{x_{bin}} \ ,$$ 
where $n(x,t)$ is the number of particles located in the bin at position $x$ and time $t$.
The density histogram is then divide by the sampling time interval $\Delta t$.

Once the average is calculated the  current is obtained with the following integration 
\begin{equation}
 {\rm J_{x}}(x,t)=-\int_{0}^{x} dx' \langle \frac{\Delta \rho^{(1)}(x',t_{s})}{\Delta  t}\rangle \ .
\end{equation}

\begin{figure}
\includegraphics[width=7cm]{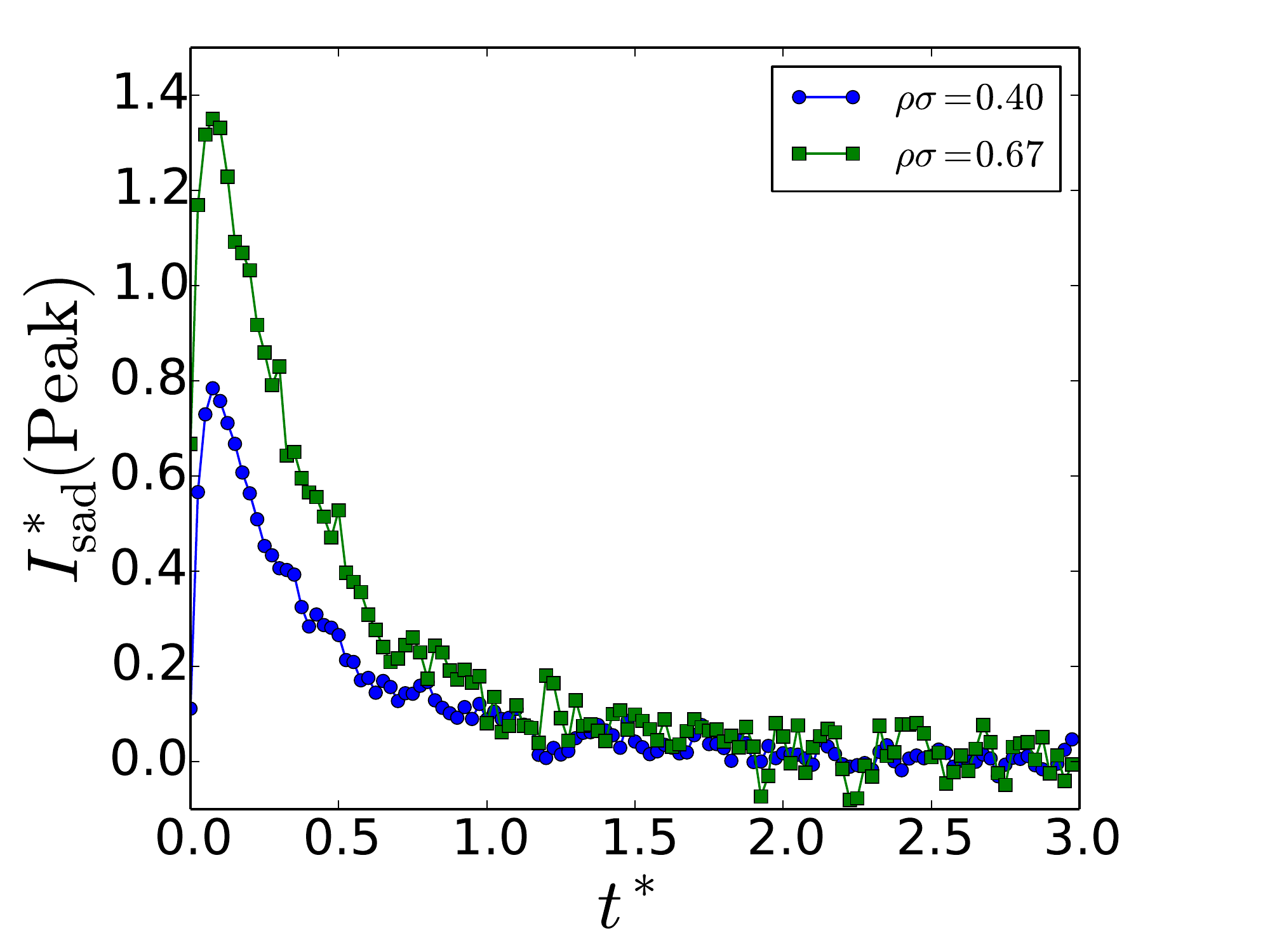}
\caption{Evolution in time of the superadiabatic force  $I^*_{\rm sad}(x_{\rm peak})=I_{\rm sad}(x_{\rm peak}) \sigma^{2}/k_{B} T$  at densities $\rho \sigma=0.4$ (green squares, $L_x=25 \sigma$) and  $\rho \sigma=0.67$ (blue circles, $L_x=15 \sigma$). }
\label{fig3}
\end{figure}

\section{Superadiabatic force}
The total pair force integral  $I(\vec r,t)=I_{\rm ad}(\vec r,t)+I_{\rm sad}(\vec r,t)$ is represented as the sum of an adiabatic term $I_{\rm ad}(\vec r,t)$, which contains all contributions that can be described by an equilibrium system, and a superadiabatic term $I_{\rm sad}(\vec r,t)$, which contains contributions that can not be reduced to an equilibrium description. 
Therefore for all equilibrium states the superadiabatic contribution vanishes.
Figure~\ref{fig3} shows the evolution in time of the superadiabatic force at  the density peak position $x_{\rm peak}$ for the system initialized  in a crystal structure. 
The superadiabatic contribution is zero for the equilibrium configurations at $t^*=0$ and $t^*\rightarrow \infty$. 
At intermediate times the curve is characterized by a  maximum at short times and by an exponential decay of the force at longer times.

\bibliography{refs_irr}
\end{document}